# Complexity-Aware Theoretical Performance Analysis of SDM MIMO Equalizers

Roya Gholamipourfard
*Optical Transmission Department*
*Nokia Bell Labs*
12 rue Jean Bart, 91300, Massy, France
roya.gholamipourfard@nokia.com

Amirhossein Ghazisaeidi
*Optical Transmission Department*
*Nokia Bell Labs*
12 rue Jean Bart, 91300, Massy, France
amirhossein.ghazisaeidi@nokia-bell-labs.com

Ruby Stella Bravo Ospina
*Optical Transmission Department*
*Nokia Bell Labs*
12 rue Jean Bart, 91300, Massy, France
ruby.bravo_ospina@nokia.com

*Abstract*—We propose a theoretical framework to compute, rapidly and accurately, the signal-to-noise ratio at the output of spatial-division multiplexing (SDM) linear MIMO equalizers with arbitrary numbers of spatial modes and filter taps and demonstrate three orders of magnitude of speed-up compared to Monte Carlo simulations.

## I. Introduction

The spatial-division multiplexing (SDM) is the way forward to support exponential data traffic growth in the years to come [1]. A massive research effort has been devoted to the study of various types of SDM: weakly and strongly coupled few-mode fibers, multi-core, and coupled-core fibers [1-5]. From the digital signal processing (DSP) perspective, the key building block is the $(2N) \times (2N) \times M$ multiple input multiple output (MIMO) linear adaptive equalizer, where $N$ is the number of spatial modes, $M$ is the number of finite-impulse response (FIR) tap coefficients, and we assume two polarizations per spatial mode, with computational complexity $O(MN^2)$, which is a major challenge in SDM system design [2]. The linear SDM channel is characterized by its overall loss, mode-dependent loss (MDL), differential-mode-delay (DMD), and random coupling [6,7]. As $N$ increases, so does the MIMO channel memory due to the DMD; therefore, $M$ should scale up proportionally to curb the equalization penalty. Moreover, the random nature of the SDM channel, requires statistical system characterization. The presence of in-line optical filtering and/or bandwidth constraints in the transmitter and receiver, further complicate the performance analysis of the MIMO equalizer by adding extra inter-symbol interference (ISI) and noise coloring. The performance analysis of SDM channels is challenging and usually is studied by time-consuming direct experiments or numerical simulations. In [8] and [9], we proposed an analytical tool for investigating the filtering penalty of 1×1 infinite-impulse response (IIR) MIMO equalizers and extended it to the 2×2 IIR case in [10] for studying polarization-dependent loss (PDL) in single-mode submarine transmission. Here, for the first time to our knowledge, we report on a complexity-aware theoretical performance analysis of $(2N) \times (2N) \times M$ FIR MIMO equalizers for optical SDM systems and present extensive validations against Monte Carlo simulations, with more than three orders of magnitude of speed-up.

## II. System Model

Fig. 1a illustrates the typical transmission scenario considered in this work. We focus on a specific target path from the transmitter (TX) to the receiver (RX), consisting of a few links supporting SDM transmission. There are optical cross-connect (OXC) nodes along the target path introducing in-line optical filtering. Links are modeled as cascades of frequency-dependent matrix channels including MDL, DMD, random coupling and in-line filtering following [6] and [7]. Fig. 1b shows the block diagram of the channel matrix model, while Fig. 1c represents the model for an OXC node with in-line filtering, and Fig. 1d illustrates the block diagram of the target path. Fig. 1e shows the total system block diagram, where for each frequency $f$, $H(f)$ is a $(2N) \times (2N)$ matrix, $W(f)$ is the covariance matrix of the colored noise, $W(f)^{-1/2}$ is the filter bank whitening the noise (note that both square root and inverse matrix operations are well-defined for positive definite matrices), and $N_0/2$ is the spectral level of the total equivalent amplified spontaneous emission (ASE) noise. Our SDM link model is entirely based on [6] and [7], and is briefly summarized in Table I.

## III. MIMO Equalizer Theoretical Model

Our model is based on the available minimum-mean-square error (MMSE) MIMO equalization theory for wireless systems [11], adapted to the optical SDM system. Assuming the oversampling factor of $s$, the vector of the received samples (*i.e.*, at MIMO input) after the $k^{th}$ channel use is denoted by $y_k$ and is of size $2Ns \times 1$. We have

$$y_k = \sum_{m=0}^{\nu} P_m \, x_{k-m} + n_k, \quad (1)$$

where $P_m$ is a matrix of size $2Ns \times 2N$, and represents the $m^{th}$ MIMO channel matrix-valued coefficient, the vector $x_{k-m}$ denotes the channel input of size $2N \times 1$ at time $k-m$, and $n_k$ is the $2Ns \times 1$ AWGN equivalent noise vector. We assume that the channel impulse response is limited to a finite time interval $[0, \nu T]$, with any nonzero amplitude outside this interval considered negligible, where $T$ is the symbol period. Assuming $Ms$ taps, fractionally spaced at $T/s$, Eq. (1) can be cast into an equivalent matrix form over blocks of length $M$ as follows

$$Y_k \triangleq \begin{bmatrix} y_k \\ y_{k-1} \\ \vdots \\ y_{k-M+1} \end{bmatrix} = \begin{bmatrix} P_0 & P_1 & \cdots & P_\nu & 0 & \cdots & 0 \\ 0 & P_0 & P_1 & \cdots & P_\nu & \cdots & 0 \\ \vdots & & \ddots & & \ddots & & \vdots \\ 0 & \cdots & 0 & P_0 & P_1 & \cdots & P_\nu \end{bmatrix} \times$$

$$\begin{bmatrix} x_k \\ x_{k-1} \\ \vdots \\ x_{k-M-\nu+1} \end{bmatrix} + \begin{bmatrix} n_k \\ n_{k-1} \\ \vdots \\ n_{k-M-\nu+1} \end{bmatrix} = P X_k + N_k, \quad (2)$$

where the matrix $P$ is of size $2NMs \times 2N(M+\nu)$. The linear MMSE MIMO FIR equalizer consists of a filter bank of discrete time-domain impulses $W$ of size $2N \times 2NMs$, which is given by the following formula

$$W = R_{xy} R_{yy}^{-1}. \quad (3)$$

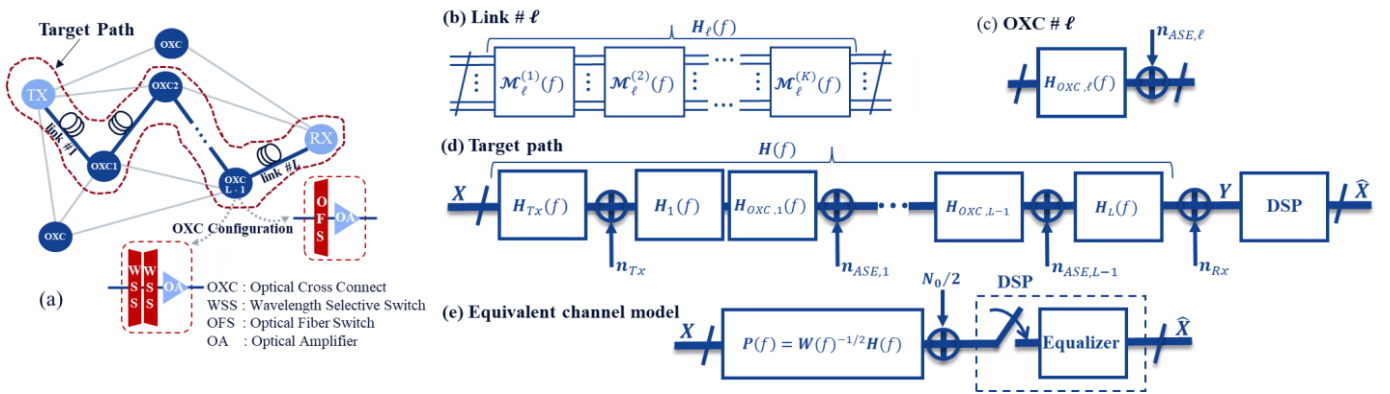

Fig.1. (a) Generalized terrestrial multi-mode optical fiber network, (b) System model of ℓth link decomposed into $K$ sections, (c) System model of ℓth node OXC, (d) Target path system model representation, (e) Equivalent digital system model.

TABLE I. SDM channel model after [6] and [7]

| | |
|---|---|
| (a) $\mathcal{M}_\ell^{(k)}(f) = \mathcal{V}_\ell^{(k)} \Lambda_\ell^{(k)}(f) \mathcal{U}_\ell^{(k)H}$    $k = 1,2,...,K$ | $\mathcal{V}_\ell^{(k)}$ and $\mathcal{U}_\ell^{(k)}$ : Unitary $(2N)\times(2N)$ random coupling matrices in the $k^{th}$ section of the $\ell^{th}$ link. Superscript $H$ stands for Hermitian conjugate. |
| (b) $\Lambda_\ell^{(k)}(f) = \mathrm{diag}\left[e^{\frac{1}{2}g_1^{(k,\ell)} - j2\pi f \tau_1^{(k,\ell)}}, ..., e^{\frac{1}{2}g_{2N}^{(k,\ell)} - j2\pi f \tau_{2N}^{(k,\ell)}}\right]$ | $g_n^{(k,\ell)}$ and $\tau_n^{(k,\ell)}$ : Gaussian random variables with standard deviations $\sigma_{mdl}$ and $\sigma_{dmd}$ per section, which are MDL loss exponent and the mode delay of the $n^{th}$ mode in the $k^{th}$ section of the $\ell^{th}$ link. |
| (c) $H_\ell(f) = \mathcal{M}_\ell^{(K)}(f) ... \mathcal{M}_\ell^{(2)}(f) \mathcal{M}_\ell^{(1)}(f)$ | $\sum_{n=1}^{2N} g_n^{(k,\ell)} = 0$ , $\sum_{n=1}^{2N} \tau_n^{(k,\ell)} = 0$. |

The input–output cross-correlation and the output auto-correlation matrices in (3) are written as

$$R_{xy} = \mathbb{E}\{x_{k-\Delta} Y_k^H\} = Z_\Delta P^H, \quad (4)$$

$$R_{yy} = \mathbb{E}\{Y_k Y_k^H\} = P P^H + s \frac{N_0}{2} I_{2NMs}, \quad (5)$$

where $\mathbb{E}\{.\}$ denotes the expectation, and $Z_\Delta$ is defined as $Z_\Delta = [0_{2N \times 2N\Delta} \ \ I_{2N} \ \ 0_{2N \times 2N(M+\nu-\Delta-1)}]$, the $I_n$ and $0_{m \times n}$ denote an $n \times n$ identity matrix and a matrix of zeros of size $m \times n$, respectively. The delay parameter $\Delta$ accounts for the channel latency. The $2N \times 1$ error vector at time $k$ is given by $E_k = x_{k-\Delta} - W Y_k$. The error autocorrelation matrix of size $2N \times 2N$ is equal to

$$R_{ee} = \mathbb{E}\{E_k E_k^H\} = I_{2N} - W R_{xy}^H$$
$$= s \frac{N_0}{2} Z_\Delta \left[ P^H P + s \frac{N_0}{2} I_{2N(M+\nu)} \right]^{-1} Z_\Delta^H. \quad (6)$$

The unbiased SNR at the $i$th output is given by

$$SNR_i = R_{ee,i}^{-1} - 1, \quad 1 \le i \le 2N \quad (7)$$

where $R_{ee,i}$ refers to the $i$th diagonal element of $R_{ee}$.

## IV. RESULTS

Our theoretical framework consists of the optical SDM system model as per Fig. 1 and Table I, along with Eqs. (1)-(7) for MIMO equalizer. In order to validate this tool, we considered single-mode and four-spatial mode cases ($N = 1$, and $N = 4$), and polarization multiplexing. For each of these two cases, we further looked at two regimes of 'low' MDL ($\sigma_{mdl} = 0.7$ dB per section) and 'high' MDL ($\sigma_{mdl} = 3.8$ dB per section), while for all four scenarios the DMD was considered 'high' (see below). We fix the AWGN noise level $N_0/2$ to -10 dB, assume root-raised cosine pulse-shape with roll-off 0.1, and no in-line filtering, and use (a), (b), (c) in Table I with $K = 50$ sections to generate ten random channel realizations per scenario. Then, for each channel realization and each scenario, we consider FIR tap counts equal to 40, 60 and 100; therefore, in total, we have 120 validation cases. For each case, once we compute the $N$ SNR values as per (7) (labelled 'theory'), and the other time conduct Monte Carlo simulations by sending waveforms and processing the received waveforms by an adaptive least mean square (LMS) equalizer as fully described in [12].

In Monte Carlo simulations, at the transmitter side, either 2 or 8 sequences (depending on the scenario) of 380,000 QPSK symbols are generated at 30 GBd. The complex constellations are fed into 0.1 roll-off factor root-raised cosine shaping filters. The shaped signals are then sent to I/Q Mach-Zehnder modulator (MZM) models for electro-optical conversion. The optical signals are then launched into the channel realizations described above using 50 sections, with each section having a length of 10 km. The channel is represented by 1,000 frequency bins spread over the simulation bandwidth. The resolution of the channel in the frequency domain is adjusted by replicating channel matrices between simulated frequency bins. The DMD standard deviation is set to 11.1 ps/√km, which corresponds to 35 ps per section length of 10 km. Nonlinear transmission effects are not simulated. After propagation, the receiver front-end model converts the received signals from the optical to the electrical domain. No phase noise has been considered for the simulations. The electric signals are down-sampled to two samples per symbol (*i.e.*, $s = 2$) and fed into the time domain equalizer (TDE). For data reception, 2×2 or 8×8 MIMO time-domain equalization is carried out by 4 or 64 finite impulse response filters updated by a fully supervised LMS algorithm [12]. For each channel realization and each scenario, we examined all $2N$ modal SNRs at the MIMO equalizer output but given the limited space in this document, in Fig. 2 we report the harmonic mean of the output MIMO SNRs per case, *i.e.*, SNR$^{-1}$ = ($\Sigma_i$ SNR$_i^{-1}$)/$2N$. This is justified as in practice, the MIMO equalizer outputs are eventually time-multiplexed into a single stream and a unique channel encoding/decoding is applied to them, but this was mainly done to simplify visualizing the results. Each subfigure corresponds to a scenario where harmonic SNR is plotted against the channel realization index. Markers represent Monte Carlo LMS simulations, while lines are the theoretical model. The solid black curve is the IIR theoretical SNR value (based on Eq. (4) in [12]). We see that theory and simulation match very well, and the penalty due to the insufficient tap counts is captured by theory. In the limit of high tap counts, both simulated and theoretical results tend towards the upper limit of IIR MMSE value. Furthermore, the actual performance fluctuations due to high MDL are perfectly captured. We benchmarked our simulations and found out for each 8-mode SNR point our theoretical model is more than 1000 times faster than the numerical simulations. Moreover, the numerical simulations should be repeated a few times for each tap count to optimize the learning coefficient in the adaptive linear equalizer, making our theory more than four orders of magnitude faster than pure Monte Carlo simulations. Another advantage of our approach is that, not only the SNRs are provided but also the actual equalizer taps are available as per (3). In Fig. 3, we consider again the previous 8-mode and high MDL scenario, but this time validate it against a new Monte Carlo simulation, where instead of the LMS, the theoretical filter bank of Eq. (3) is used to simulate the receiver. Finally, in Fig. 4, we illustrate a more advanced application of our tool, where 4 links, 4 spatial modes and high MDL scenario are assumed, and we also consider 4 in-line 2-nd order super-Gaussian optical filters with 3 dB bandwidth 15 GHz. These 4 filters are once placed all at the transmitter side (TX), a second time all at the receiver side (RX) and the third time, evenly distributed at the end of each link. Fig. 4 illustrates the harmonic SNR vs. AWGN noise variance for a fixed random channel realization, but three filtering scenarios and various number of taps, showing complexity-aware linear channel equalization for future optical SDM networks.

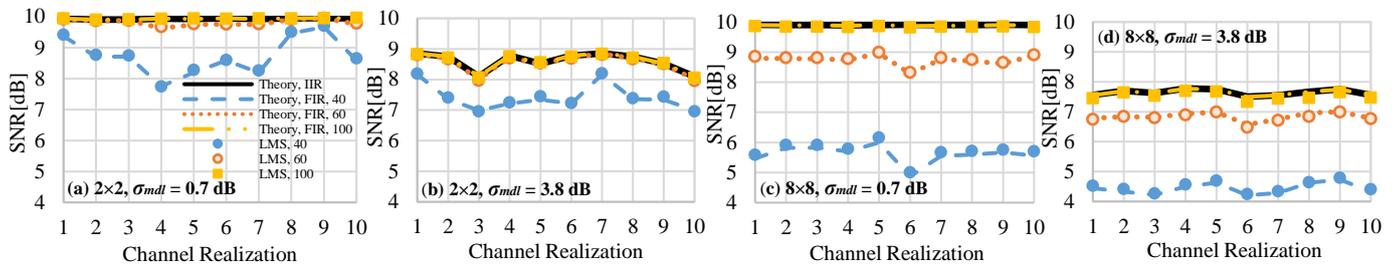

Fig. 2. Comparison of harmonic SNR between analytical model predictions (lines) and simulations based on LMS (markers) for a 30 GBd QPSK signal transmitted through a 2 and an 8-mode fiber, for $\sigma_{mdl} = 0.7, 3.8$ dB, $\sigma_{dmd} = 35$ ps per section, and $N_0/2 = -10$ dB.

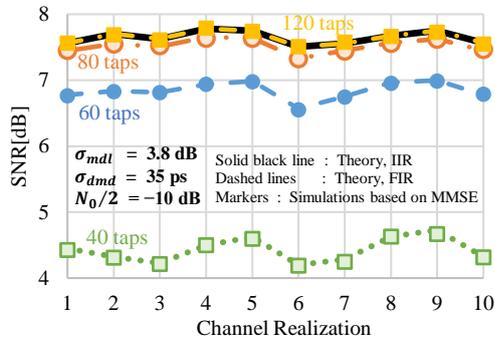

Fig. 3. Comparison of system performance with IIR and FIR equalizers for different number of taps within an 8-mode fiber.

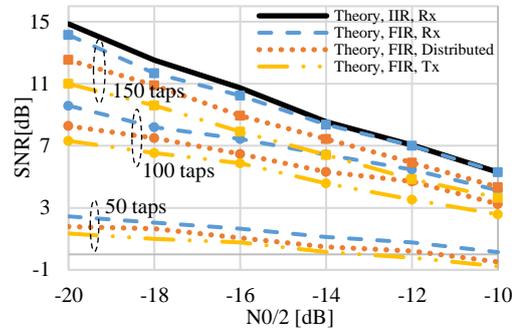

Fig. 4. Harmonic SNR vs. AWGN noise variance for three filtering scenarios and various number of taps within an 8-mode fiber.

## V. CONCLUSIONS

We presented, for the first time to our knowledge, a comprehensive analytical tool to analyze the performance of MIMO equalizers for arbitrary number of modes and taps and validated it against extensive Monte Carlo simulations. We demonstrated up to 4 orders of magnitude of speed up compared to detailed Monte Carlo simulations with the same accuracy in performance prediction, which makes our proposal an interesting design and optimization tool for future optical spatial-division multiplexing networks.